\documentclass[final,5p,twocolumn]{elsarticle}

\usepackage{amssymb}

\usepackage{bibentry}
\usepackage{amsmath}
\usepackage{multicol}
\usepackage{fancyhdr}
\usepackage{subcaption}
\usepackage{lineno}

\usepackage[labelfont=bf,justification=raggedright,singlelinecheck=false]{caption}
\journal{Nuclear Physics B}

\begin{document}

\begin{frontmatter}


\title{Application of a PWFA to an X-ray FEL}


\author{Yasmine Israeli, Jorge Vieira,  Sven  Reiche, Marco Pedrozzi  and Patric Muggli}

\address{Max Planck Institute for Physics,\\ Instituto Superior T\' ecnico,\\ Paul Scherrer Institute }

\begin{abstract}
There is a growing demand for X-ray Free-electron lasers (FELs) in various science fields, in particular for  those with short pulses, larger photon fluxes and shorter wavelengths.
The level of X-ray power and the pulse energy depend on the amount of  electron bunch energy. Increasing the latter will increase the power of the radiating X-rays.\\
 Using numerical simulations we explore  the possibility of using a plasma wakefield accelerator (PWFA) scheme to increase the electron beam energy of an existing FEL facility without significantly increasing the accelerator length. In this paper we use parameters of the SwissFEL beam.
The simulations are carried out in 2D cylindrical symmetry using the code OSIRIS. Initial results show an energy gain of $\sim$2 GeV after propagation of 0.5 m in the plasma with a relative energy spread of $\sim$1\%.
\end{abstract}

\end{frontmatter}
\section{Introduction}
Free Electron Lasers (FELs) provide very intense and tightly focused X-ray beams. These X-rays can be used to map the atomic structure of materials, including bio-molecules and nanometer scale structures. 
The laser power and the radiation wavelength are determined by the energy and brightness of the electron bunches.
 In this paper, we present an FEL scheme which includes a plasma wakefield accelerator (PWFA) after the linear accelerator (linac).\\ 
 
PWFAs can provide accelerating gradients up to 100 GV/m, orders of magnitudes higher than gradients that can be produced by conventional radio frequency linacs. By adding a PWFA after the linac, we may be able to double the electron energy in a  much shorter distance than that of the linac and  potentially generate a pulse with higher energy and a shorter wavelength.\\

One of the promising future FELs, SwissFEL, is  being constructed by Paul Scherrer Institute (PSI). The beam parameters in this study have been chosen from within the range of possible operating modes of SwissFEL. 
\section{Swiss FEL Beam Parameters }
The SwissFEL baseline design seeks to provide a wavelength range from 0.1 nm to 7 nm. The undulator design,  chosen for the minimum wavelength (0.1 nm), has a  period length, $\lambda_{U}$, of 15 mm and an undulator parameter, $K$, equal to 1.2 \cite{CDR}. 
The radiation wavelength is given by 
\begin{linenomath}
\begin{align}
\lambda_{ph}=\dfrac{\lambda_{U}}{2\gamma^2}\left( 1+\dfrac{K^2}{2} \right),
\end{align}
\end{linenomath}
where $\gamma$ is the electrons energy in units of the rest energy $m_{e}c^2$.
Therefore, for this operation the energy of the electron beam is 5.8 GeV.\\

The accelerator facility enables different operation modes.  The beam charge can typically range between 10 pC and 200 pC and the bunch size ($\sigma_{z}, \sigma_{r}$),  from 10 $\mu$m to  20 $\mu$m.  Beams with 200 pC are characterized by normalized projected emittance $\epsilon_{N}=0.43$ mm.mrad and a 350 keV energy spread.  
The saturation power can be estimated via $ P_{sat}\sim \rho P_{beam}$, with $P_{beam}=\gamma mc^2\cdot I_e/e$.\\
$\rho$ is the Pierce parameter, defined as \cite{Huang}:
\begin{linenomath}
\begin{align}
\rho= \left[  \dfrac{1}{16}\frac{I_{e}}{I_{A}} \left( \dfrac{K[JJ]\lambda_{U}}{\sigma_{r}2\pi}\right) ^2 \dfrac{1}{\gamma^3}\right] ^{1/3},
\end{align}\\
\end{linenomath}
where $I_{e}$ is the electron beam peak current, $I_{A}$ is Alfven current ($\approx17$ kA) and the Bessel function factor $[JJ]$
is equal to $J_{0}\left(K^2/(4+2K^2)\right)-J_{1}\left(K^2/(4+2K^2)\right)$.\\ 

By increasing the electron beam energy to 13.6 GeV with $\lambda_U$=30 mm and K=2.75\footnote{Using $K=0.934\cdot B[T]\cdot\lambda_U$[cm],  $B$ is the peak magnetic field of the undulator on the axis.}, we can increase the saturation power by a factor of 2.5. In addition, the lasing requirements: $\dfrac{\sigma_{\gamma}}{\gamma}<\rho$ and $\dfrac{•\epsilon_{N}}{•\gamma}\leqslant \dfrac{•\lambda_{ph}}{•4\pi}$, where $\sigma_{\gamma}$ is the local intrinsic energy spread, are better fulfilled with higher beam energies.
Using a PWFA might make this energy jump possible within a compact acceleration distance.

\section{The PWFA Scheme}
A PWFA fires a driving bunch into a plasma and uses the resulting oscillation of plasma electrons to accelerate a witness bunch. An effective acceleration  depends on the bunch parameters (charge $Q$, longitudinal size $\sigma_{z}$ and transverse size $\sigma_{r}$) as well as on the plasma density $n_{pe}$. The most important is the drive bunch longitudinal size, which should be in the same  order of the plasma wavelength.  \\

\begin{figure}
\centering
\includegraphics[width=8cm]{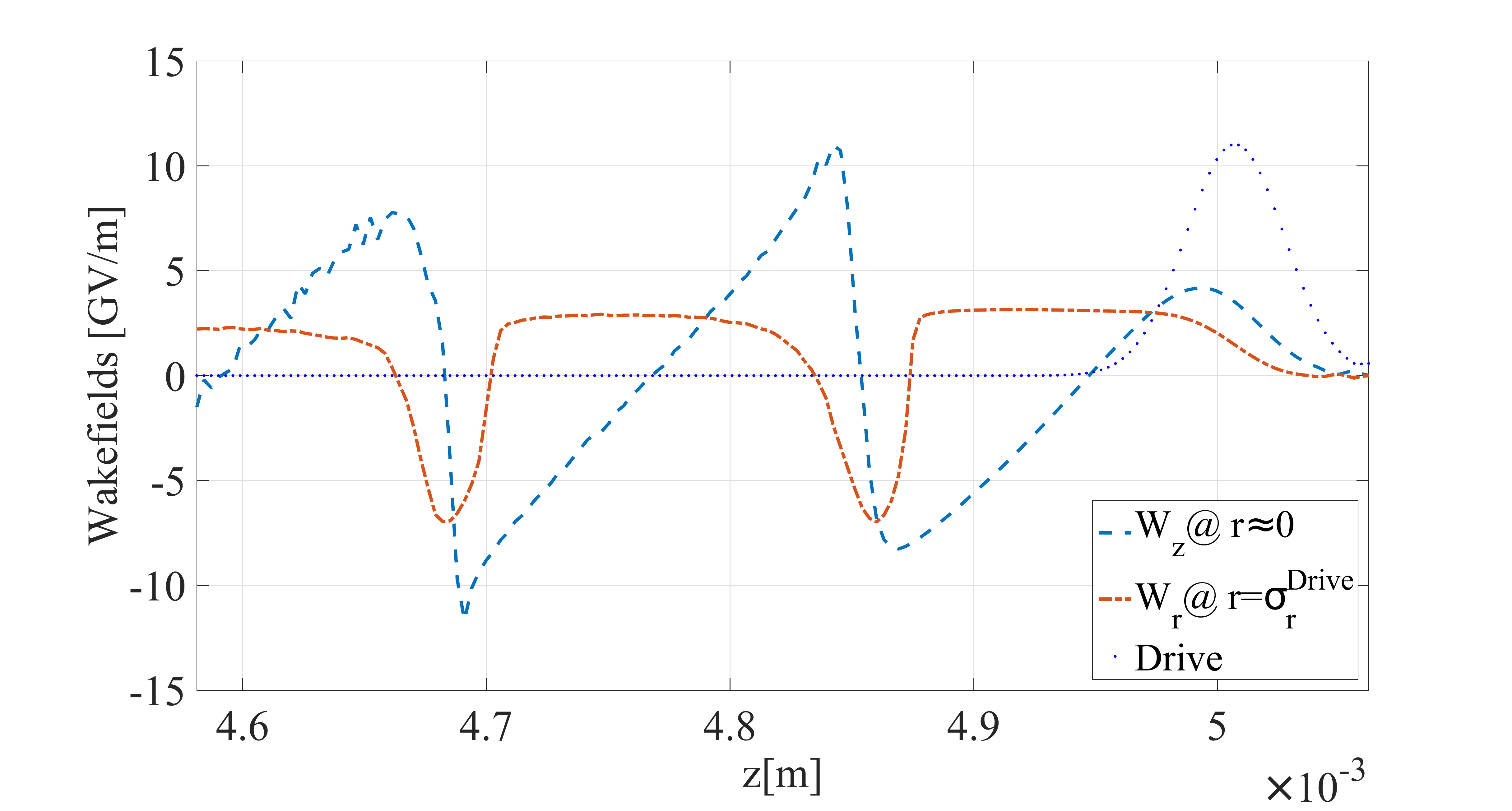}
\caption{Lineouts of the longitudinal wakefield $W_{z}$ on axis (--) and the transverse wakefield $W_r$  at $\sigma_r^{Drive}$ (-.) after 5 mm propagation in the plasma. Result from 2D cylindrical OSIRIS simulation with  $n_{b}/n_{pe}=$2.245. The simulation parameters are summarized in Table \ref{tab:parameters}. }
 \label{fig:wf}
\end{figure}

This study focuses on the non-linear regime  (or the bubble regime), where the drive bunch density $n_{b}$ exceeds  the plasma density.  
In the non-linear regime the transverse wakefield, $W_{r}$, inside each bubble is independent of the propagation direction $z$ (Figure \ref{fig:wf}) and linear with the radius $r$ \cite{NL}. Therefore, the bunch  experiences a constant focusing force inside the bubble, minimizing the emittance growth through the propagation. 
In addition, within each bubble the longitudinal wakefield, $W_{z}$, is independent of the radius and thus equally accelerates  particles with the same longitudinal position. However, the accelerating field varies linearly with the longitudinal position, leading to a correlated  energy spread for an accelerated witness bunch. By means of simulations we investigate the  loading of the wakefield by the witness bunch in order to reduce the final energy spread.\\

\begin{figure}
\centering
\includegraphics[width=8cm]{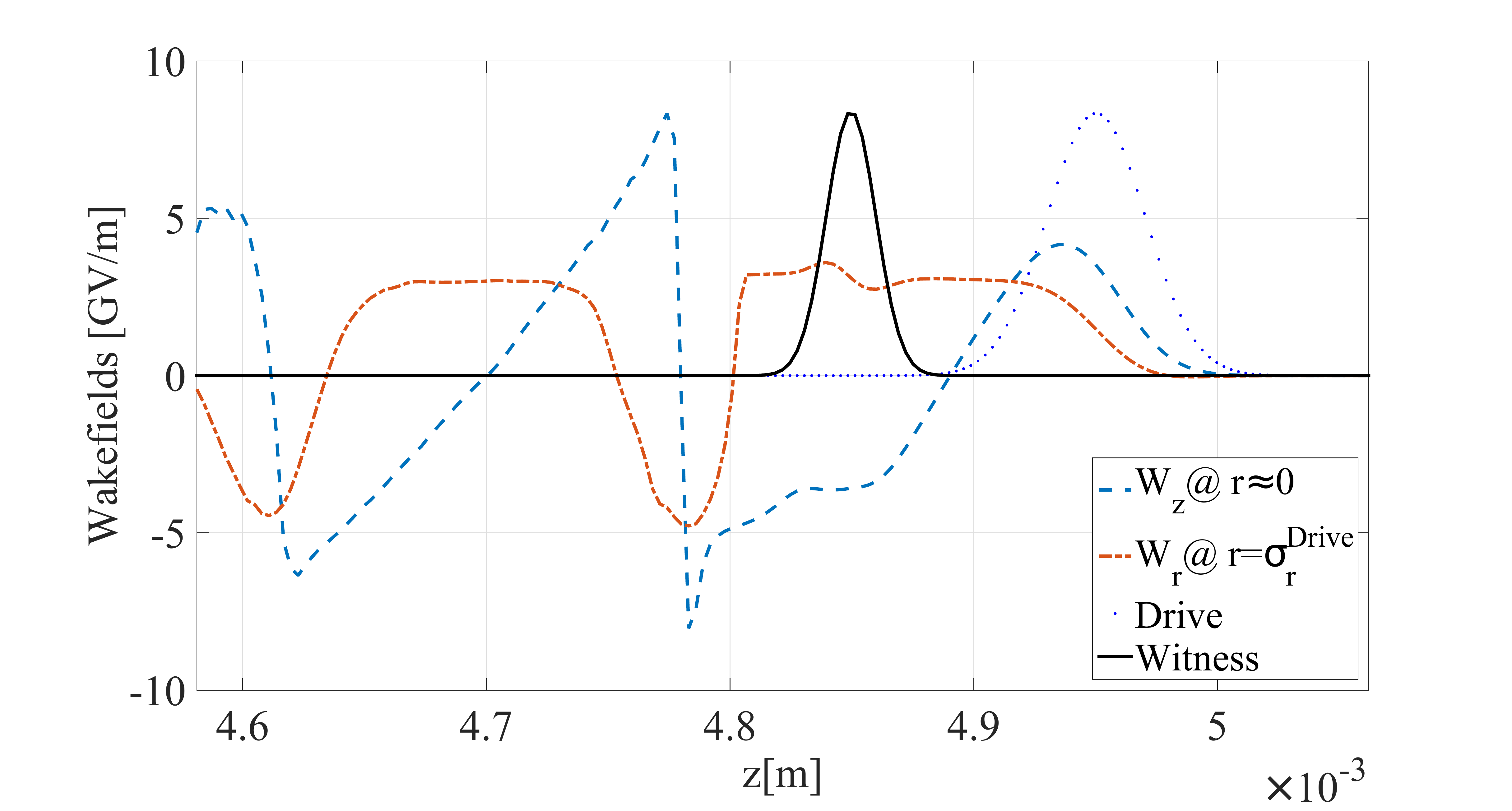}
 \caption{Lineouts of the longitudinal wakefield $W_{z}$ on axis (--) and the transverse wakefield $W_r$ at $\sigma_r^{Drive}$ (-.) after 5 mm  propagation in the plasma. The drive and witness bunches are marked by dotted (.) and solid lines respectively. Result from 2D cylindrical OSIRIS simulation with  $n_{b}/n_{pe}=$2.245. The simulation parameters are summarized in Table \ref{tab:parameters}.  }
 \label{fig:bl}
\end{figure}

\begin{figure*}
\begin{center}
\begin{minipage}[]{0.45\linewidth}
\centering
\includegraphics[width=8cm,height=5cm]{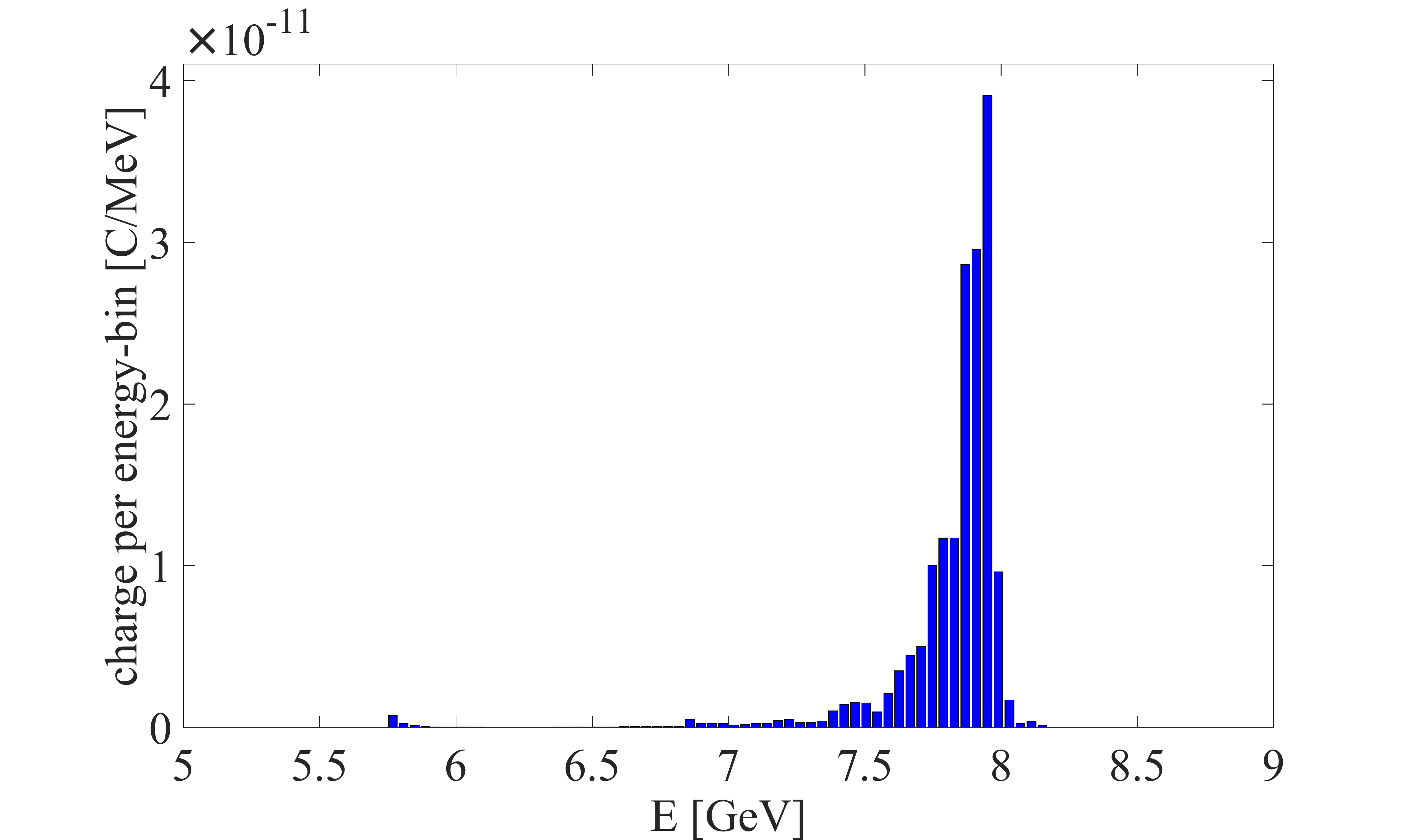} 
\end{minipage}
\hfill%
\begin{minipage}[h]{0.45\linewidth}
\centering
\includegraphics[width=8cm,height=5cm]{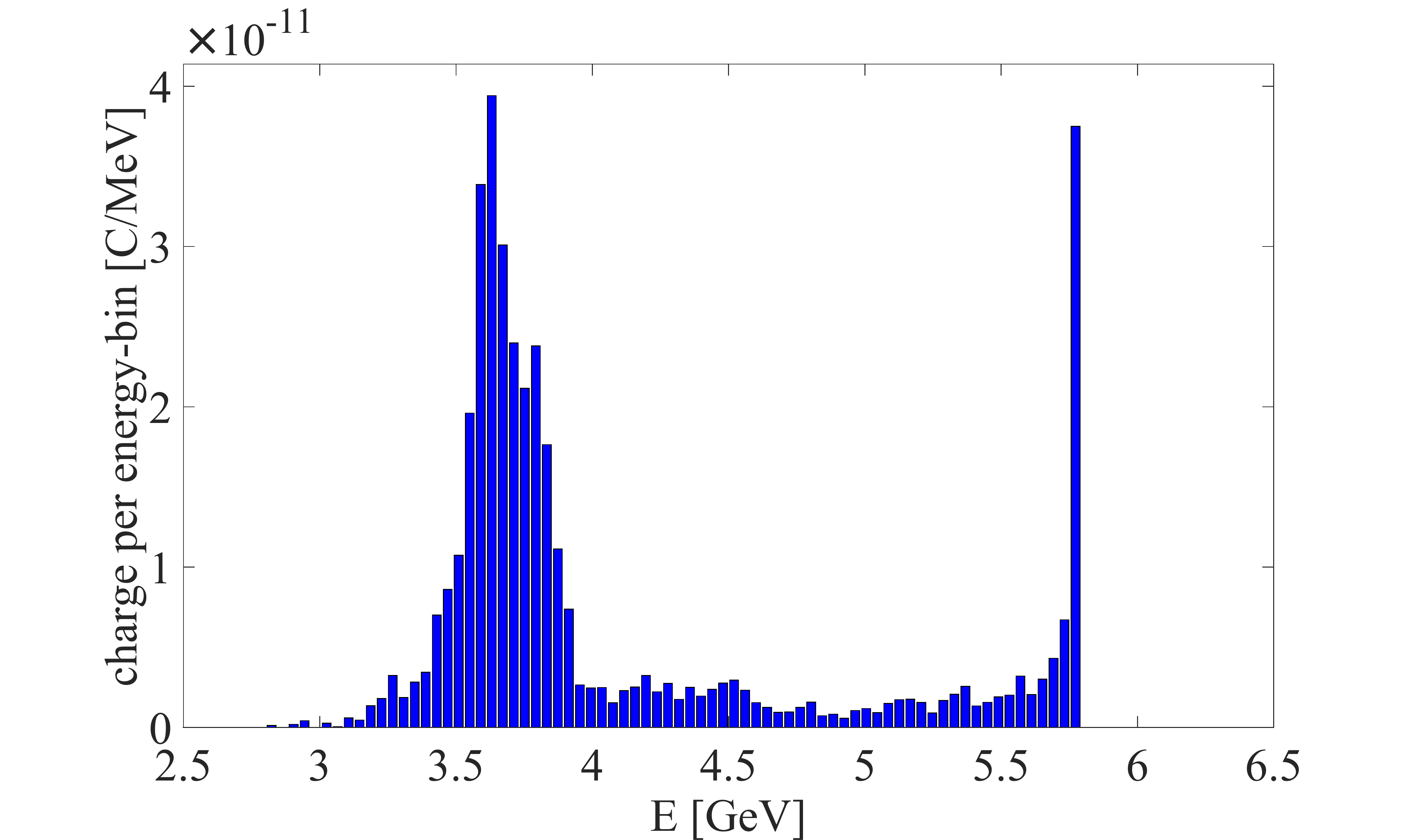} 
\end{minipage}
\hfill%
\caption{Energy distributions of the witness (left) and drive (right) bunches after propagation of 0.5 m inside the plasma. The witness bunch distribution has a maximum value at 7.94 GeV and FWHM of 80.32 MeV (includes 57\% of particles of the initial beam). The drive bunch mean value is 4.13 GeV with FWHM of 2.2 GeV (52\% of the particles). The bin width is 40 MeV.}
\label{fig:es}
\end{center}
\end{figure*}

For the PWFA design, we use OSIRIS \cite{OSIRIS} code to perform numerical simulations with 2D cylindrical symmetry.
Our preliminary PWFA scheme includes a plasma with  $n_{pe}=3.53\cdot10^{16}$ cm$^{-3}$ and a 400 pC drive bunch with $\sigma_{z}^{Drive}$=20 $\mu$m and  $\sigma_{r}^{Drive}$=10 $\mu$m. 
The selection of the witness bunch parameters is based on the beam loading  principles \cite{BeamLoading} in order to minimize the energy-spread. In the current design  a 170 pC witness bunch with transverse and longitudinal lengths of 10 $\mu$m is placed 5$\cdot\sigma_{z}^{Drive}$ from the drive bunch. For simplicity, the bunches are assumed to be mono-energetic with an energy of 5.8 GeV at injection in the plasma.  The parameters for this scenario are summarized in Table \ref{tab:parameters}.\\

 Figure \ref{fig:bl} presents the longitudinal and transverse wakefields after propagation of 5 mm of the drive and witness bunches in the plasma . We can see that the accelerating gradient $G$ reaches about 3.5 GeV/m and the field is relatively constant in the witness bunch.
From this initial accelerating gradient, we can estimate the average energy gain of the witness bunch after $L_p$=0.5 m to be $G\cdot L_{P}=$1.75 GeV.\\
\begin{table}
\centering
\begin{tabular}{|c|c|}
\hline 
$n_{pe}$ & 3.53$\cdot10^{16}$ cm$^{-3}$ \\
\hline 
peaks separation & 100 $\mu$m  \\ 
\hline
initial energy & 5.8 GeV \\
\hline
$\sigma_{z}^{Drive}$ & $20$ $\mu$m \\
\hline 
$\sigma_{r}^{Drive}$ & $10$ $\mu$m \\
\hline
$Q^{Drive}$ & 400 pC \\
\hline
$\sigma_{z}^{Witness}$&  $10$ $\mu$m \\
\hline 
$\sigma_{r}^{Witness}$ &  $10$ $\mu$m \\
\hline 
 $Q^{Witness}$ & 170 pC \\
\hline
\end{tabular} 
\caption{Parameters  of the preliminary PWFA scheme.}
\label{tab:parameters}
\end{table}

 Figure \ref{fig:es} presents the energy distributions of the drive and the witness bunches at 0.5 m. 
The witness bunch distribution has a maximum value at 7.94 GeV and a full width at half maximum (FWHM) of 80.32 MeV with 57\% of the  bunch particles. 
The mean value of the drive bunch is 4.13 GeV and the  FWHM is 2.1 GeV. Accordingly, the witness bunch  gained $2.14$ GeV while the drive bunch lost $1.67$ GeV. The relative energy spread of the witness bunch is about 1\%, yet does not satisfy the SwissFEL requirements.

\section{Conclusions}
FEL saturation power  depends on the energy of the electron bunches. In the case of SwissFEL, reaching an energy level of 13.6 GeV will increase the power by a factor of 2.5. A PWFA scheme with SwissFEL bunch parameters and $n_{pe}=3.53\cdot10^{16}$ cm$^{-3}$  can reach an accelerated field with a multi GeV/m scale. In this first study we show an energy gain of $\sim2$ GeV in 0.5 m for the witness bunch. Consequently, we can assume a doubling of the beam energy in  very few meters. \\

We loaded the plasma wave and minimized the energy spread of the witness bunch to 80.3 MeV. However SwissFEL operation requires an energy spread of 350 keV. An useful approach would be to investigate  particles with an energy range of $\pm$175 keV around the maximum distribution  value. This would insure a suitable energy spread,  while reducing the applicable charge for lasing.\\

We note here that loading of the longitudinal field leads to a modification of the transverse field in the witness bunch as can be seen in Figure \ref{fig:bl}.
Future studies will aim at reducing the witness bunch energy spread as well as at minimizing potential emittance growth due to "loading" of the transverse wakefields.

\section{Acknowledgment}
The authors would like to acknowledge the OSIRIS Consortium, consisting of UCLA and IST (Lisbon, Portugal) for the use of OSIRIS, for providing access to the OSIRIS framework.


\bibliographystyle{unsrt} 
\bibliography{EAAC}
\end{document}